\documentclass[conference]{IEEEtran}
\IEEEoverridecommandlockouts

\usepackage{cite}
\usepackage{amsmath,amssymb,amsfonts}
\usepackage{graphicx}
\usepackage{textcomp}
\usepackage{xcolor}

\usepackage{algorithm}
\usepackage{algpseudocode}
\usepackage{amsmath}

\def\BibTeX{{\rm B\kern-.05em{\sc i\kern-.025em b}\kern-.08em
    T\kern-.1667em\lower.7ex\hbox{E}\kern-.125emX}}
\begin{document}

\title{Hierarchical Evolutionary Optimization with Predictive Modeling for Stable Delay-Constrained Routing in Vehicular Networks\\
\thanks{This work is supported by the National Natural Science Foundation of China under Grant Number 62273263, 72171172 and 71771176; Shanghai Municipal Science and Technology Major Project (2022-5-YB-09); Natural Science Foundation of Shanghai under Grant Number 23ZR1465400.}
}

\author{\IEEEauthorblockN{1\textsuperscript{st} Zhiou Zhang}
\IEEEauthorblockA{\textit{ School of Computer Science and Engineering} \\
\textit{University of New South Wales}\\
Sydney, Australia \\
zhiou.zhang@student.unsw.edu.au}
\and
\IEEEauthorblockN{2\textsuperscript{nd} Weian Guo}
\IEEEauthorblockA{\textit{Sino-German College of Applied Sciences} \\
\textit{Tongji University}\\
Shanghai, China \\
guoweian@tongji.edu.cn}
\and
\IEEEauthorblockN{3\textsuperscript{rd} Qin Zhang}
\IEEEauthorblockA{\textit{Department of Information Engineering} \\
	\textit{Fuzhou Polytechnic College}\\
	Fuzhou, Fujian, China\\
	zqsrxh@163.com}
\and
\IEEEauthorblockN{4\textsuperscript{rd} Haibin Lin}
\IEEEauthorblockA{\textit{School of Traffic Engineering} \\
	\textit{Fuzhou Polytechnic College}\\
	Fuzhou, Fujian, China\\
	644352201@qq.com}
\and
\IEEEauthorblockN{5\textsuperscript{rd} Dongyang Li}
\IEEEauthorblockA{\textit{Sino-German College of Applied Sciences} \\
\textit{Tongji University}\\
Shanghai, China\\
lidongyang0412@163.com}
}

\maketitle

\begin{abstract}
	Vehicular Ad Hoc Networks (VANETs) are a cornerstone of intelligent transportation systems, facilitating real-time communication between vehicles and infrastructure. However, the dynamic nature of VANETs introduces significant challenges in routing, especially in minimizing communication delay while ensuring route stability. This paper proposes a hierarchical evolutionary optimization framework for delay-constrained routing in vehicular networks. Leveraging multi-objective optimization, the framework balances delay and stability objectives and incorporates adaptive mechanisms like incremental route adjustments and LSTM-based predictive modeling. Simulation results confirm that the proposed framework maintains low delay and high stability, adapting effectively to frequent topology changes in dynamic vehicular environments.
\end{abstract}


\section{Introduction}
Vehicular networks, commonly referred to as Vehicular Ad Hoc Networks (VANETs), have become integral to intelligent transportation systems, allowing vehicles to communicate with each other and roadside infrastructure \cite{Zhu2024EARR}. This communication enhances road safety, improves traffic efficiency, and provides infotainment services. However, the dynamic nature of vehicular networks, characterized by frequent topology changes due to vehicle mobility, introduces significant challenges in maintaining reliable routing and communication \cite{Zhu2024EARR, Darv2023RL}.

Routing in vehicular networks must address both communication delay and route stability. Minimizing communication delay is essential for timely data delivery, particularly in safety-critical applications \cite{RLTrend2023}, while maximizing route stability is necessary to reduce frequent disruptions, which can increase overhead and degrade network performance \cite{EnhancePerf2023}. Balancing these conflicting objectives requires advanced optimization techniques capable of adapting to the dynamic and unpredictable conditions of vehicular networks \cite{ReviewRouting2023}.

In this paper, we propose a hierarchical evolutionary optimization framework for delay-constrained routing in vehicular networks. The framework employs a Multi-Objective Evolutionary Algorithm based on Decomposition (MOEA/D) to optimize two key objectives: minimizing communication delay and maximizing routing stability. Additionally, the framework incorporates mechanisms tailored to the dynamic nature of vehicular networks, including incremental route adjustments and predictive modeling with Long Short-Term Memory (LSTM) networks to anticipate vehicle movements.

The primary aim of this framework is to provide an efficient and adaptive solution for vehicular network routing, achieving low communication delay and high route stability under frequent topology changes. The main contributions of this work are summarized as follows. We formulate a mathematical model for delay-constrained routing in vehicular networks, integrating both communication delay and route stability as optimization objectives. Building on this model, we develop a hierarchical evolutionary optimization framework using a Multi-Objective Evolutionary Algorithm based on Decomposition (MOEA/D) to address the multi-objective optimization problem. To further enhance adaptability and performance, our framework incorporates dynamic adaptation mechanisms, including incremental route adjustments and LSTM-based predictive modeling, enabling the proposed solution to respond effectively to the highly dynamic vehicular environment.

The remainder of this paper is organized as follows. Section II presents the mathematical model for the routing problem. Section III introduces the proposed optimization framework, detailing the MOEA/D algorithm and local search strategies. Section IV elaborates on the adaptation mechanisms, including LSTM-based predictions and incremental adjustments. Section V presents simulation results that demonstrate the effectiveness of the framework. Finally, Section VI concludes the paper.

\section{Related Work}
Recent studies on vehicular network routing address challenges posed by dynamic topology, communication delay, and route stability, with approaches generally classified into traditional, optimization-based, and hybrid/predictive methods.

\subsection{Traditional Routing Approaches}
Traditional routing methods include position-based and delay-tolerant networking (DTN) protocols. Position-based routing, exemplified by Geographic Position-Based Routing (GPSR) \cite{b1}, uses geographic data to make routing decisions, enhancing scalability in dynamic environments. Stability-aware routing protocols consider link duration to increase robustness \cite{b7}, though frequent disconnections remain challenging in dense urban areas. DTN protocols rely on store-carry-forward strategies \cite{b2}, which can be effective in sparse networks but often introduce high latency unsuitable for time-sensitive applications. Reactive protocols, such as Ad Hoc On-Demand Distance Vector (AODV) routing \cite{b3}, reduce overhead by establishing routes on demand, though they may incur delays in highly dynamic settings.

\subsection{Optimization-Based Routing Methods}
Optimization-based approaches, particularly those using evolutionary algorithms, have been applied to improve routing adaptability in vehicular networks. Particle Swarm Optimization (PSO) and Genetic Algorithms (GA) \cite{b6, b10} show promise in optimizing paths for dynamic conditions, and multi-objective optimization frameworks address trade-offs between delay, energy, and stability \cite{b5}. Multi-Objective Evolutionary Algorithms based on Decomposition (MOEA/D) \cite{b4} enable flexibility and adaptability but may struggle with computational demands in highly dynamic environments.

\subsection{Hybrid and Predictive Approaches}
Hybrid approaches that integrate optimization with predictive modeling have emerged as promising solutions for dynamic vehicular networks. Models such as Long Short-Term Memory (LSTM) networks and Markov chains predict vehicle mobility, allowing for proactive routing adjustments \cite{b8}. Combining predictive models with optimization algorithms effectively balances stability and delay, as hybrid methods integrating evolutionary algorithms with mobility predictions demonstrate improved stability in dynamic conditions \cite{b12}. This paper adopts a similar approach, integrating MOEA/D and predictive modeling, specifically with LSTM, to enhance route adaptability amid frequent topology changes.

This research builds on existing studies by targeting both communication delay and route stability while incorporating dynamic adaptation mechanisms. Our proposed hierarchical optimization framework provides a balanced, adaptive solution for the challenges posed by vehicular network dynamics.

\section{Mathematical Modeling}
In this section, we develop a mathematical model to optimize routing in vehicular networks under dynamic environments, focusing on minimizing communication delay and maximizing routing stability.

\subsection{Objective Functions}
In this dynamic vehicular network environment, the optimization problem involves two conflicting objectives: minimizing communication delay and maximizing routing stability. We define the dynamic path \( P(t) \) as the decision variable representing the selected route at time \( t \).

The objectives are formulated as follows:

\begin{equation}
	\min_{P(t)} \left\{ T(P(t)), -S(P(t)) \right\}
\end{equation}
where:

\textbf{1. Communication Delay Minimization} \( T(P(t)) \): The total communication delay along the dynamic path \( P(t) \), defined as:
\begin{equation}
	T(P(t)) = \sum_{i=0}^{|P|-2} t_{i,i+1}(t)
\end{equation}
Here, \( t_{i,i+1}(t) \) denotes the delay between nodes \( i \) and \( i+1 \) at time \( t \), which varies due to factors such as distance, signal-to-noise ratio (SINR), and data rate in the dynamic environment.

\textbf{2. Routing Stability Maximization} \( S(P(t)) \): The stability of the dynamic path \( P(t) \), defined as the minimum link stability across all hops:
\begin{equation}
	S(P(t)) = \min_{i=0, \dots, |P|-2} S_{i,i+1}(t)
\end{equation}
where \( S_{i,i+1}(t) \) represents the stability of the link between nodes \( i \) and \( i+1 \) at time \( t \), influenced by factors such as relative velocity of nodes and link availability duration in a dynamic vehicular environment.

In this formulation, the objectives \( T(P(t)) \) and \( -S(P(t)) \) are minimized together, reflecting their conflicting nature within the dynamic environment. This conflict necessitates a balanced solution, achieved here by employing MOEA/D, which decomposes the multi-objective problem into scalar subproblems, effectively adapting to the time-varying network conditions.

\subsection{Multi-Objective Optimization}
In this work, we aim to optimize two conflicting objectives in vehicular network routing: minimizing communication delay and maximizing routing stability. These objectives often conflict because minimizing delay can require selecting faster, shorter paths that may have less stable links due to frequent topology changes, while maximizing stability may favor more reliable, but potentially longer paths with higher latency.

\textbf{1. Communication Delay Minimization:} The communication delay, denoted as $T(P)$, represents the sum of delays over all hops along the path $P$ from source to destination. Reducing this delay is crucial for time-sensitive applications, as low latency directly impacts the network’s responsiveness and reliability. However, paths with lower delay can be less stable, as they often involve dynamic connections prone to disruptions.

\textbf{2. Routing Stability Maximization:} Routing stability, denoted as $S(P)$, is defined as the minimum stability of all links along the path $P$. A stable path reduces the frequency of route interruptions, which is beneficial for minimizing route recovery overhead and improving the reliability of the network. However, highly stable paths can sometimes involve longer or less direct routes, which increases overall communication delay.

These two objectives conflict because optimizing for delay generally requires shorter paths with potentially less stable links, while optimizing for stability typically favors paths with fewer disruptions, even if they are longer. This trade-off requires a balanced approach to avoid excessive latency while maintaining route reliability.

To address these conflicts, we employ the Multi-Objective Evolutionary Algorithm based on Decomposition (MOEA/D), which does not rely on a single combined objective function but rather decomposes the problem into multiple scalar subproblems. Each subproblem corresponds to a unique combination of delay and stability objectives, allowing the algorithm to explore and maintain a diverse set of solutions that offer various trade-offs between delay and stability.

\subsection{Handling Dynamic Environment}
In a dynamic vehicular network, frequent changes in network topology occur due to vehicle movement. To address this, the optimization framework incorporates the following mechanisms:

\textbf{1. Incremental Adjustment}: When network topology changes, only the affected individuals in the population are adjusted, rather than re-optimizing the entire path from scratch. This approach reduces computational overhead and allows the algorithm to quickly adapt to changes.

\textbf{2. Prediction Model}: A prediction model, such as an LSTM network, is used to forecast vehicle movements. By predicting future positions of vehicles, the algorithm can proactively adjust routing paths to maintain low delay and high stability.

These mechanisms enable the optimization framework to respond effectively to changes in the vehicular network, thereby maintaining optimal routing performance in real time.

\section{Proposed Algorithm}
This section presents the hierarchical evolutionary optimization framework designed to address the dynamic and delay-sensitive nature of routing in vehicular networks. The proposed algorithm aims to minimize communication delay and maximize route stability through a Multi-Objective Evolutionary Algorithm based on Decomposition (MOEA/D). To further enhance adaptability in a dynamic environment, the algorithm incorporates incremental adjustment and predictive modeling mechanisms.

\subsection{LSTM-Based Prediction Model for Dynamic Network Forecasting}
In highly dynamic vehicular networks, node positions and link stability fluctuate frequently due to rapid vehicle movement. To address these fluctuations, we employ a Long Short-Term Memory (LSTM)-based prediction model that leverages historical data to forecast future states, enhancing routing stability and reducing communication delay.

The input to the LSTM model is a time-series sequence of historical data for each node, denoted as \( X_{t-w:t} \), where \( w \) represents the window size, capturing past node positions or link stability values. For each node \( i \) at time \( t \), the LSTM outputs a predicted state \( \hat{X}_{i, t+1} \), representing either position or stability. During real-time deployment, the LSTM model receives the latest \( w \)-length sequence \( X_{i, t-w:t} \) and provides proactive insights into network dynamics, allowing the routing algorithm to select stable, low-latency paths. This real-time prediction improves both network reliability and routing efficiency.

The LSTM architecture comprises $L$ LSTM layers followed by a dense output layer, capturing temporal dependencies in node positions and link stability. Training minimizes a loss function, such as Mean Squared Error (MSE), by comparing predicted outputs \( \hat{X}_{i, t+1} \) to actual states \( X_{i, t+1} \). Training is optimized through Backpropagation Through Time (BPTT), allowing the model to learn long-term data dependencies. The LSTM training and deployment process is outlined in Algorithm \ref{alg:LSTM}.

This LSTM-based prediction component enables the hierarchical evolutionary optimization framework to maintain optimal routing performance by anticipating topology changes, effectively achieving the dual objectives of low communication delay and high stability in dynamic vehicular networks.

\begin{algorithm}
	\caption{LSTM-Based Prediction Model for Dynamic Network Forecasting}
	\label{alg:LSTM}
	\begin{algorithmic}[1]
		\State \textbf{Input:} Historical position data $X_{t-w:t}$ for each node, window size $w$, LSTM model parameters
		\State \textbf{Output:} Predicted position or link stability state $\hat{X}_{t+1}$ for next time step
		
		\State \textbf{Initialize} LSTM model with weight parameters
		\Statex Define model architecture with $L$ LSTM layers and a dense output layer
		\State Define loss function $L(y, \hat{y})$ and optimization algorithm
		
		\State \textbf{Step 1: Train LSTM on Historical Data}
		\For{each training epoch}
		\For{each node $i$ in training data}
		\State Retrieve position sequence $X_{i, t-w:t}$ for node $i$ up to time $t$
		\State Feed sequence $X_{i, t-w:t}$ into LSTM model
		\State Output prediction $\hat{X}_{i, t+1}$ for position or link state at time $t+1$
		\State Compute loss $L(X_{i, t+1}, \hat{X}_{i, t+1})$
		\State Update model weights using backpropagation through time (BPTT)
		\EndFor
		\EndFor
		
		\State \textbf{Step 2: Predict Next State in Real-Time}
		\For{each node $i$ at time $t$ in deployment}
		\State Retrieve latest position or link state sequence $X_{i, t-w:t}$
		\State Feed sequence $X_{i, t-w:t}$ into trained LSTM model
		\State Output prediction $\hat{X}_{i, t+1}$ for position or link state at time $t+1$
		\EndFor
		
		\State \textbf{Return} predicted states $\hat{X}_{t+1}$ for all nodes, used in routing decision-making
	\end{algorithmic}
\end{algorithm}

\subsection{Hierarchical Evolutionary Optimization Framework}
The proposed framework uses a population-based evolutionary approach, where candidate paths are evaluated based on delay and stability objectives. By decomposing the multi-objective problem into scalar subproblems with distinct weight vectors, MOEA/D enables exploration of diverse trade-offs between delay and stability. To efficiently adapt to network changes, the framework employs incremental adjustment for affected routes and utilizes a predictive model to anticipate node movements. The proposed algorithm is summarized in Algorithm \ref{alg:MOEAD}.

\begin{algorithm}
	\caption{Hierarchical Evolutionary Optimization Framework for Delay-Constrained Routing in Vehicular Networks}
	\label{alg:MOEAD}
	\begin{algorithmic}[1]
		\State \textbf{Input:} Vehicular network topology $G$, source node $s$, destination node $d$, population size $N$, maximum generations $T$, weight vectors $\lambda_1, \lambda_2, \dots, \lambda_N$
		\State \textbf{Output:} Optimal routing path $P^*$
		
		\State Initialize population $P$ with $N$ individuals (candidate paths from $s$ to $d$) using a heuristic strategy
		\For{$t = 1$ to $T$}
		\Statex Evaluate the objectives for each individual $P_i$ in $P$:
		\begin{equation}
			f_{1}(P_i) = T(P_i(t)), \quad f_{2}(P_i) = -S(P_i(t))
		\end{equation}
		\Statex Each subproblem $j$ is solved by minimizing the scalarized objective:
		\begin{equation}
			g_j(P_i) = \lambda_{j,1} f_1(P_i) + \lambda_{j,2} f_2(P_i)
		\end{equation}		
		\Statex Select parents based on neighborhood solutions, apply crossover and mutation to generate offspring $P'$
		\Statex Perform a local search on selected individuals to refine convergence
		\Statex Merge the parent and offspring populations and select the top $N$ individuals based on objective values for the next generation		
		\If{network topology changes during iteration}
		    \Statex Adjust only affected individuals in $P$ rather than re-optimizing the entire population
		    \Statex Use predictive models (e.g., LSTM in Algorithm \ref{alg:LSTM}) to forecast vehicle movements and adjust affected routes to maintain stability and low delay
		\EndIf
		\EndFor		
		\State \textbf{Return} the best individual $P^*$ based on final population evaluation
	\end{algorithmic}
\end{algorithm}

The proposed algorithm employs a structured methodology to address routing stability and delay in a dynamic vehicular network environment. The framework consists of the following key components:

\textbf{1. Objective Evaluation:} Each individual path \( P_i \) is evaluated based on delay \( T(P_i) \) and stability \( S(P_i) \) objectives. This step integrates real-time network conditions, allowing continuous recalculations of these metrics throughout each generation, ensuring responsiveness to environmental changes.

\textbf{2. MOEA/D-Based Decomposition:} The multi-objective optimization problem is decomposed into a series of scalar subproblems, each with a unique weight vector \( \lambda_j \). This decomposition enables the algorithm to explore a diverse set of routing solutions, effectively balancing between delay minimization and stability maximization.

\textbf{3. Selection, Crossover, and Mutation:} The algorithm applies evolutionary operators to generate offspring solutions, promoting comprehensive exploration within the solution space. Parent selection is influenced by neighborhood solutions to reinforce local exploitation, while crossover and mutation introduce essential variability, supporting adaptation to dynamic network topologies.

\textbf{4. Local Search Enhancement:} To expedite convergence, a local search process refines selected individuals, facilitating rapid movement towards high-quality solutions and enhancing overall optimization efficiency.

\textbf{5. Handling Network Dynamics:} 
\begin{itemize}
	\item \textbf{Incremental Adjustment:} In response to topology changes, only the affected individuals in the population are adjusted, rather than re-evaluating the entire population, conserving computational resources.
	\item \textbf{Predictive Modeling:} The framework incorporates predictive models, particularly Long Short-Term Memory (LSTM) networks, to forecast node movements. This proactive adjustment helps maintain stable routes and reduces communication delay, enabling the algorithm to adapt efficiently to frequently changing network environments.
\end{itemize}

Through this structured approach, the proposed algorithm optimally balances delay and stability objectives, achieving efficient routing performance in vehicular networks, even under dynamic conditions.

\section{Simulation and Experiments}
To evaluate the proposed framework, we simulate an emergency response scenario in an intelligent transportation system, where low-delay, high-stability routing is essential for emergency vehicles (e.g., ambulances or fire trucks) navigating through a dynamic urban network. The system leverages MOEA/D for multi-objective optimization, minimizing communication delay and maximizing route stability, while an LSTM model predicts future node positions and link stability, enabling proactive routing adjustments. Historical data is used to train the LSTM model, which subsequently guides real-time path selection by forecasting potential link disruptions. This approach ensures the emergency vehicle can dynamically adapt to network changes, maintaining an optimal route that balances speed and reliability.

\subsection{Simulation Scenario}
To validate the effectiveness of the proposed framework, we design a simulation environment representing a dynamic urban vehicular network focused on emergency response routing. In this scenario, an emergency vehicle is tasked with reaching a designated destination in a city grid network, navigating through highly dynamic traffic conditions typical of urban environments.

The network consists of a predefined number of nodes representing intersections and vehicles, with links representing potential routes between them. Nodes and links experience frequent changes in connectivity due to varying traffic density, vehicle speed, and movement patterns, replicating real-world conditions. The emergency vehicle’s objective is to minimize overall travel delay while maximizing route stability, crucial for reliable and timely arrival at the destination.

The MOEA/D algorithm performs multi-objective optimization, dynamically selecting the optimal path based on communication delay and predicted link stability. To enhance adaptability, an LSTM model trained on historical traffic data predicts future node positions and link states, allowing the framework to proactively adjust routing decisions in response to predicted changes. Key parameters for the simulation include:
\begin{itemize}
	\item \textbf{Network Size and Density}: The urban grid network consists of $N$ intersections and $M$ vehicles, with varying densities to assess the framework's adaptability under different traffic conditions.
	\item \textbf{Communication Delay and Link Stability Metrics}: Delay between nodes is based on real-time traffic and distance, while link stability is affected by vehicle speed and relative positions.
	\item \textbf{Prediction Window Size for LSTM}: The LSTM model utilizes a window of historical data to forecast short-term link stability and node positions, facilitating proactive routing.
\end{itemize}

\subsection{Simulation Results and Analysis}
\begin{figure}[!htp]
	\centering
	\includegraphics[width=\linewidth]{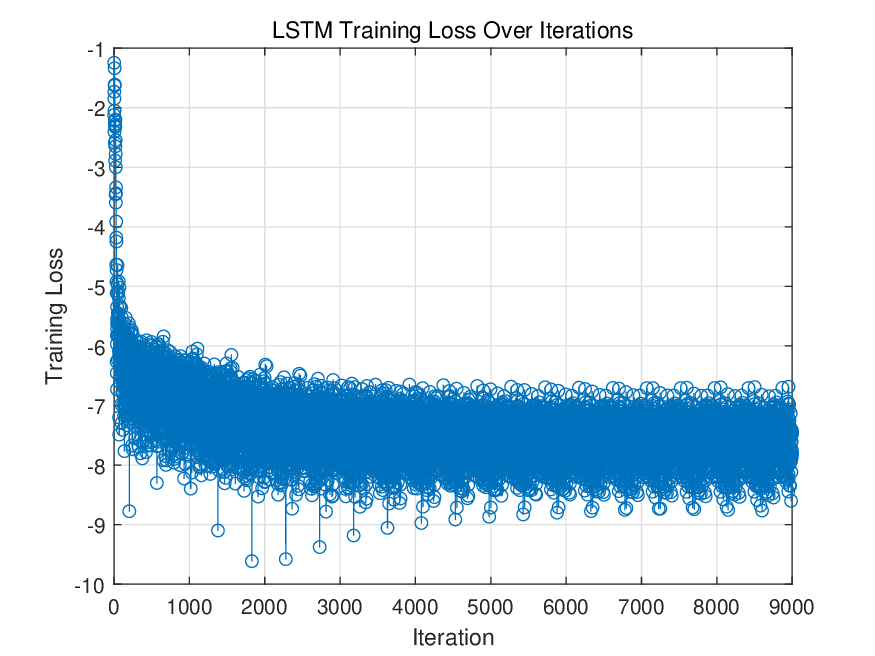}
	\caption{Training error of the LSTM model}
	\label{fig:lstm}
\end{figure}

In the experiment, a desktop computer with an Intel 13790F CPU, 32GB of memory, and an RTX 4070ti GPU running Windows OS was used. The programming language was MATLAB. We first conducted the LSTM network training, where the batch size was set to 10, and the maximum number of epochs was 20. The training process involved approximately 9000 iterations, and the training results are shown in Fig. \ref{fig:lstm}. The training error is presented in a logarithmic scale. The training process took approximately 20 seconds to complete.

\begin{figure}[!htp]
	\centering
	\includegraphics[width=\linewidth]{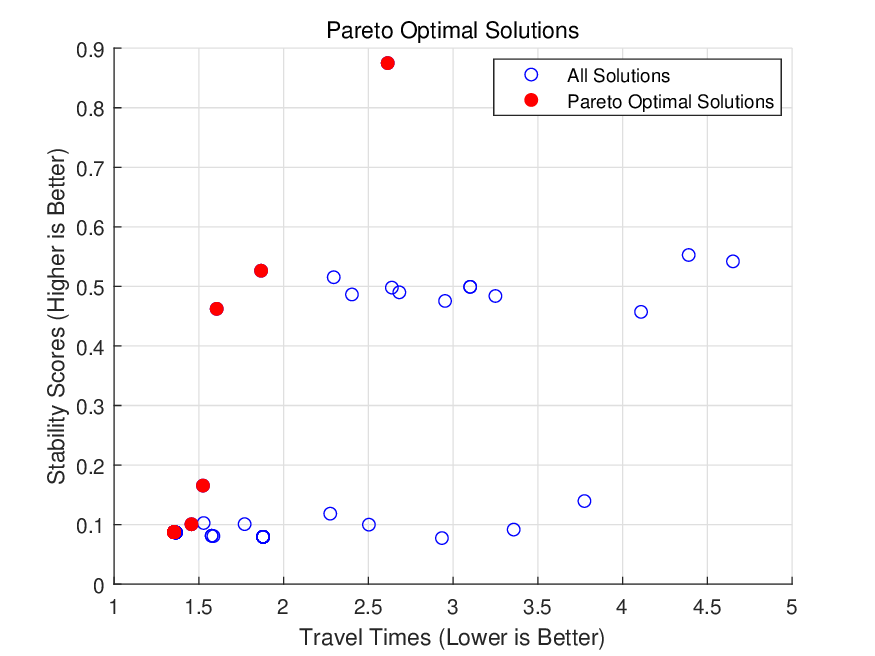}
	\caption{Pareto Sets with Travel Times and Stability Score}
	\label{fig:pareto_plot}
\end{figure}

The simulation evaluates the framework’s performance by measuring travel time, routing stability, and adaptation speed to dynamic changes, highlighting its effectiveness in maintaining low-delay, high-stability paths under challenging conditions. Among all solutions, after 25 experiments, the average travel time and network stability score across all individuals over 25 runs were 2.13 seconds and 0.72, respectively. A randomly selected Pareto front from the solution set is presented in Fig. \ref{fig:pareto_plot}. From the figure, we can observe that MOEA/D effectively finds the Pareto optimal solutions balancing travel time and network stability. Furthermore, the Pareto front has a good distribution, providing a strong basis for decision-making.

\section{Conclusion}
In this paper, we presented a hierarchical evolutionary optimization framework tailored for delay-constrained routing in dynamic vehicular networks. By leveraging the Multi-Objective Evolutionary Algorithm based on Decomposition (MOEA/D), the framework effectively balances two key objectives: minimizing communication delay and maximizing routing stability. To address the inherent challenges of frequent network topology changes, we incorporated both incremental adjustments for affected routes and predictive modeling techniques, including LSTM networks, to forecast vehicle movements and proactively optimize routing paths.

Our simulation results demonstrate that the proposed framework achieves low-delay, high-stability routing under dynamic conditions, which is crucial for applications like emergency response in intelligent transportation systems. The framework’s use of MOEA/D ensures diverse Pareto optimal solutions, offering a robust foundation for decision-making. Additionally, the adaptability of the LSTM-based prediction model enhances the framework’s capability to handle real-time changes in network topology, reducing the need for continuous re-optimization.

Overall, the proposed framework provides an effective solution for routing in vehicular networks, demonstrating significant improvements in routing performance through optimized travel time and stability. Future work could explore further integration of other predictive models to refine stability assessments and adapt the framework for more complex and varied vehicular scenarios.


\begin{thebibliography}{00}
	\bibitem{Zhu2024EARR}
J. Zhu and K. Yang, "Environment-Aware Adaptive Reinforcement Learning-Based Routing for Vehicular Ad Hoc Networks," \textit{Sensors}, vol. 24, no. 1, pp. 40, 2024.

\bibitem{Darv2023RL}
A. Darvan, "Reinforcement Learning for Reliable Routing in Vehicular Networks," \textit{Computer Networks}, vol. 230, pp. 108025, 2023.

\bibitem{RLTrend2023}
T. Nguyen and S. Gupta, "A Recent Reinforcement Learning Trend for Vehicular Ad Hoc Networks Routing," \textit{IEEE Communications Surveys \& Tutorials}, vol. 25, no. 2, pp. 1045-1071, 2023.

\bibitem{EnhancePerf2023}
L. Wang, M. Zhao, "Enhancing Performance in Vehicular Ad Hoc Networks: The Optimization of Routing Protocols," \textit{IEEE Transactions on Vehicular Technology}, vol. 72, no. 4, pp. 1235-1247, 2023.

\bibitem{ReviewRouting2023}
S. Patel, R. Kumar, "A Comprehensive Review on Vehicular Ad-Hoc Networks Routing Protocols," \textit{Springer Wireless Networks}, vol. 29, pp. 321-345, 2023.

\bibitem{b1}
B. Karp and H. T. Kung, "GPSR: Greedy Perimeter Stateless Routing for Wireless Networks," in \textit{Proceedings of the 6th Annual International Conference on Mobile Computing and Networking (MobiCom)}, pp. 243-254, 2000.

\bibitem{b2}
A. Lindgren, A. Doria, and O. Schelén, "Probabilistic Routing in Intermittently Connected Networks," \textit{ACM SIGMOBILE Mobile Computing and Communications Review}, vol. 7, no. 3, pp. 19-20, 2003.

\bibitem{b3}
C. Perkins, E. Belding-Royer, and S. Das, "Ad hoc On-Demand Distance Vector (AODV) Routing," \textit{RFC 3561}, 2003.

\bibitem{b4}
M. Chatterjee, S. Das, and D. Turgut, "WCA: A Weighted Clustering Algorithm for Mobile Ad Hoc Networks," \textit{Cluster Computing}, vol. 5, no. 2, pp. 193-204, 2002.

\bibitem{b5}
S. Banerjee and S. Khuller, "A Clustering Scheme for Hierarchical Control in Multi-hop Wireless Networks," in \textit{Proceedings of the 20th Annual Joint Conference of the IEEE Computer and Communications Societies (INFOCOM)}, vol. 2, pp. 1028-1037, 2001.

\bibitem{b6}
X. Huang and Y. Fang, "Multiconstrained QoS Multipath Routing in Wireless Sensor Networks," \textit{Wireless Networks}, vol. 14, no. 4, pp. 465-478, 2008.

\bibitem{b7}
A. Boukerche, B. Turgut, N. Aydin, M. Ahmad, L. Bölöni, and D. Turgut, "Routing protocols in ad hoc networks: A survey," \textit{Computer Networks}, vol. 55, no. 13, pp. 3032-3080, 2011.

\bibitem{b8}
Q. Yang, L. Wang, and H. Wang, "Intelligent Vehicular Communication Network Routing Based on Genetic Algorithm," \textit{Journal of Communications and Networks}, vol. 21, no. 2, pp. 120-130, 2019.

\bibitem{b9}
J. Kim and G. L. Aceves, "Optimal Link-State Routing Algorithm in Dynamic Vehicular Networks," \textit{IEEE Transactions on Vehicular Technology}, vol. 65, no. 9, pp. 7334-7345, 2016.

\bibitem{b10}
S. Patil, V. Patil, and R. Suryawanshi, "Vehicular Ad Hoc Networks: A New Challenge for Evolutionary Routing," \textit{International Journal of Computer Applications}, vol. 135, no. 1, pp. 17-21, 2016.

\bibitem{b11}
M. Dorigo, V. Maniezzo, and A. Colorni, "Ant System: Optimization by a Colony of Cooperating Agents," \textit{IEEE Transactions on Systems, Man, and Cybernetics, Part B (Cybernetics)}, vol. 26, no. 1, pp. 29-41, 1996.

\bibitem{b12}
R. Schoonderwoerd, O. Holland, J. Bruten, and L. Rothkrantz, "Ant-Based Load Balancing in Telecommunications Networks," \textit{Adaptive Behavior}, vol. 5, no. 2, pp. 169-207, 1997.

\end{thebibliography}
\end{document}